# An efficient polynomial time approximation scheme for load balancing on uniformly related machines

Leah Epstein[*]     Asaf Levin[†]

**Abstract**

We consider basic problems of non-preemptive scheduling on uniformly related machines. For a given schedule, defined by a partition of the jobs into $m$ subsets corresponding to the $m$ machines, $C_i$ denotes the completion time of machine $i$. Our goal is to find a schedule which minimizes or maximizes $\sum_{i=1}^{m} C_i^p$ for a fixed value of $p$ such that $0 < p < \infty$. For $p > 1$ the minimization problem is equivalent to the well-known problem of minimizing the $\ell_p$ norm of the vector of the completion times of the machines, and for $0 < p < 1$ the maximization problem is of interest. Our main result is an efficient polynomial time approximation scheme (EPTAS) for each one of these problems. Our schemes use a non-standard application of the so-called shifting technique. We focus on the work (total size of jobs) assigned to each machine and introduce intervals of forbidden work. These intervals are defined so that the resulting effect on the goal function is sufficiently small. This allows the partition of the problem into sub-problems (with subsets of machines and jobs) whose solutions are combined into the final solution using dynamic programming. Our results are the first EPTAS's for this natural class of load balancing problems.

## 1 Introduction

We consider non-preemptive scheduling problems on $m$ uniformly related machines. In such problems, we are given a set of jobs $\{1, 2, \ldots, n\}$, where each job $j$ has a positive size $p_j$. The jobs need to be partitioned into $m$ subsets $S_1, \ldots, S_m$, with $S_i$ being the subset of jobs assigned to machine $i$. We let $s_i$ denote the speed of machine $i$, and the processing of a job $j$ takes $\frac{p_j}{s_i}$ time units if $j$ is assigned to machine $i$. For such a solution (also known as a schedule), we let $C_i = \frac{\sum_{j \in S_i} p_j}{s_i}$ be the *completion time* of machine $i$. The *work* (also called the weight) of machine $i$ is $W_i = \sum_{j \in S_i} p_j = C_i \cdot s_i$, that is, the total size of the jobs which are assigned to $i$. The makespan of the schedule is $\max_i C_i$, and the optimization problem of finding a schedule which minimizes the makespan is well-studied (see e.g. [20, 19, 23, 24, 26]). The problem of finding a schedule which maximizes $\min_i C_i$ is the well-known Santa Claus problem on uniformly related machines (see e.g. [18, 30, 2, 5, 15, 8]). Both these problems are concerned with the optimization of the extremum values of the set $\{C_1, \ldots, C_m\}$.

Motivated by minimizing average latency in storage allocation applications (rather than worst-case latency), researchers have suggested to study the optimization goal of minimizing the $\ell_2$ norm (and the goal of minimizing the $\ell_p$ norm for $p > 1$) of the vector of completion times of the machines (see e.g. [12, 11, 28, 4, 3]). It was stated more recently by Bansal and Pruhs [7] that: "*The standard way to compromise between optimizing for the average and optimizing for the worst case is to optimize the $\ell_p$ norm, generally for something like $p = 2$ or $p = 3$.*" An additional perspective of using the $\ell_p$ norm as an objective function has arisen recently in algorithmic game theory [9]. Note that the minimization of the $\ell_p$ norm is equivalent to minimizing the sum of the $p$-th powers of the completion times of machines.

---
[*]Department of Mathematics, University of Haifa, 31905 Haifa, Israel. lea@math.haifa.ac.il.
[†]Faculty of Industrial Engineering and Management, The Technion, 32000 Haifa, Israel. levinas@ie.technion.ac.il.

Thus, we consider objective functions in which the entire vector $\mathcal{C} = (C_1, \ldots C_m)$ affects the value of the objective function. Our class of objective functions includes the minimization of the sum of the $p$-th powers of the completion times of machines which is equivalent to the minimization of the $\ell_p$ norm of $\mathcal{C}$. More precisely, given a fixed real (finite) number $p$ such that $0 < p < \infty$, we consider the problem of minimizing $\sum_{i=1}^{m} C_i^p$ and the problem of maximizing $\sum_{i=1}^{m} C_i^p$. The minimization problem for $p \leq 1$ is trivially solved by placing all the jobs on one of the fastest machines. Therefore, we consider the minimization problem only for values of $p$ such that $p > 1$. Similarly, the maximization problem is trivially solved for $p \geq 1$ by placing all the jobs on one of the slowest machines. Hence, we consider the maximization problem only for values of $p$ such that $p < 1$.

An $\mathcal{R}$-approximation algorithm for a minimization problem is a polynomial time algorithm which always finds a feasible solution of cost at most $\mathcal{R}$ times the cost of an optimal solution. An $\mathcal{R}$-approximation algorithm for a maximization problem is a polynomial time algorithm which always finds a feasible solution of value at least $\frac{1}{\mathcal{R}}$ times the value of an optimal solution (we use the convention of approximation ratios greater than 1 for maximization problems). The infimum value of $\mathcal{R}$ for which an algorithm is an $\mathcal{R}$-approximation is called the approximation ratio or the performance guarantee of the algorithm. A polynomial time approximation scheme (PTAS) is a family of approximation algorithms such that the family has a $(1 + \varepsilon)$-approximation algorithm for any $\varepsilon > 0$. An efficient polynomial time approximation scheme (EPTAS) is a PTAS whose time complexity is of the form $f(\frac{1}{\varepsilon}) \cdot poly(n)$ where $f$ is some (not necessarily polynomial) function and $poly(n)$ is a polynomial of the length of the (binary) encoding of the input. Motivated by this definition of polynomial time complexity, we say that an algorithm (for some problem) has polynomial time complexity if its time complexity is of the form $f(\frac{1}{\varepsilon}) \cdot poly(n)$. Note that whereas a PTAS may have time complexity of the form $n^{g(\frac{1}{\varepsilon})}$, where $g$ is for example linear or even exponential, this cannot be the case for an EPTAS. The notion of an EPTAS is modern and find its roots in the FPT (fixed parameter tractable) literature (see [10, 13, 17, 29]).

Our main result is a class of EPTAS's for minimizing $\sum_{i=1}^{m} C_i^p$ for any fixed value of $p > 1$, and for the problem of maximizing $\sum_{i=1}^{m} C_i^p$ for any fixed positive value of $p < 1$. Note that these problems are known to be strongly NP-hard even for identical machines (via the standard reduction from the 3-PARTITION problem) and therefore our results are the best possible. Our results are the first EPTAS's for these important load balancing problems on uniformly related machines.

The running time of an EPTAS (and of a PTAS) is expected to be polynomial in the number of jobs as well as in the number of machines. For a fixed (constant) number of machines, load balancing problems typically have a fully polynomial time approximation schemes (FPTAS's, which are EPTAS's where $f$ is polynomial) [25, 6, 14, 16].

We next review the previous PTAS and EPTAS results for an arbitrary (non-constant) number of uniformly related machines and the special case of identical machines (where all machines have unit speed). It was shown by Hochbaum and Shmoys that the makespan minimization problem has a PTAS for identical machines [23] and for uniformly related machines [24]. It was noted in [21] that the PTAS of [23] for identical machines can be converted into an EPTAS by using integer program in fixed dimension instead of dynamic programming. Recently, Jansen [26] was able to solve the long-standing open problem of establishing an EPTAS for the makespan minimization problem on uniformly related machines. The Santa Claus problem is also known to have a PTAS and an EPTAS for identical machines [30, 2]. For uniformly related machines a PTAS is known [5, 15].

The problems studied here are known to have an EPTAS on identical machines [1, 2], and a PTAS on uniformly related machines [15]. The existence of an EPTAS for these problems on uniformly related machines was stated as an open problem by [15]. This open problem is resolved in our work.

**Outline.** Our EPTAS's have the following structure. First, we sort the machines in a non-decreasing order of their weight in an optimal solution (according to either non-increasing or non-decreasing speed). We note that some machines may get a zero weight; we guess their number and remove those machines from the instance. We round the processing times of the jobs and the speeds of the machines, so that the number of possible values is reduced sufficiently, and so that all job sizes are integer multiples of a small value.

Next, we observe that we can extend the EPTAS for identical machines to the case where we are guaranteed that in an optimal solution the ratio between the maximum work of any machine and the minimum work of any machine is bounded. We show that in this case the speed ratio is bounded as well. We extend this EPTAS further to allow some total size of jobs to remain unscheduled. This will be our building block in the design of the EPTAS for the general case.

To reduce the general case into a series of sub-problems of the former type, we create gaps between the set of allowed weights of machines. For that, we apply the so-called shifting technique [22] in an original way. Afterwards, we apply dynamic programming to determine the series of sub-problems, that is, the intervals of machines whose weights come from each interval of allowed weights. The EPTAS for the special case is used as a black box in this dynamic programming, where unscheduled jobs of one sub-problem are scheduled later by another sub-problem. Omitted proofs can be found in the Appendix.

## 2 Preliminaries

In this paper we consider the sum of the $p$-th powers of a vector rather than the $(\frac{1}{p})$-th power of this value. Note that since $p$ is a fixed constant, our results apply also for this last alternative measure (which is the $\ell_p$ norm for the case $p > 1$). Throughout the paper, for a solution $\mathcal{A}$ we denote by $\mathcal{A}$ both the solution and the value of the objective function for this solution.

When we consider the maximization problem, we sometimes allow the algorithm to avoid assigning some of the jobs. It is clear that adding these jobs arbitrarily to the schedule can only improve the solution. Hence, if we can bound the total value of the solution which assigns a subset of the jobs, after adding the unscheduled jobs (to create a complete solution), we get (at least) the same performance guarantee.

Let $\varepsilon$ be a small constant such that $0 < \varepsilon < \frac{1}{2}$ and $\frac{1}{\varepsilon}$ is an integer. Epstein and Sgall [15] observed the following claim.

**Claim 1** *Let $i_1$ and $i_2$ be a pair of machines such that $s_{i_1} < s_{i_2}$, that is, $i_2$ is faster than $i_1$. Consider the minimization problem when $p > 1$, then any optimal solution satisfies $W_{i_1} \leq W_{i_2}$. Consider the maximization problem when $p < 1$, then any optimal solution satisfies $W_{i_1} \geq W_{i_2}$.*

Motivated by the above claim we will sort the machines according to their weights. That is, when we consider the minimization problem we will assume that $s_1 \leq s_2 \leq \cdots \leq s_m$, whereas when we consider the maximization problem we will assume that $s_1 \geq s_2 \geq \cdots \geq s_m$. In this way, machines of lower indices should get smaller weight than machines with higher indices (or equal weight). We next consider a pair of machines $i_1$ and $i_2$ such that $s_{i_1}$ is significantly smaller than $s_{i_2}$. We know that in the minimization problem $W_{i_1} \leq W_{i_2}$ and in the maximization problem $W_{i_1} \geq W_{i_2}$. Our next goal is to strengthen these bounds. Let $\delta$ be such that $0 < \delta \leq \varepsilon$.

**Lemma 2** *Consider the minimization problem ($p > 1$), and a pair of machines $i < i'$. There exists a function $\alpha(\delta) = \frac{\delta}{2}$ such that if $s_i \leq \alpha(\delta) \cdot s_{i'}$ then in any optimal solution $W_i \leq \delta W_{i'}$.*

**Lemma 3** *Consider the maximization problem ($p < 1$), and a pair of machines $i < i'$. There exists a function $\alpha(\delta) = ((1+\delta)^p - 1)^{1/p}$ such that if $s_{i'} \leq \alpha(\delta) \cdot s_i$ then in any optimal solution $W_i \leq \delta W_{i'}$.*

Note that $\alpha(\delta) \leq \delta \leq \varepsilon$. This is clear for the minimization problem, and for the maximization problem it holds because $(1+\delta)^p \leq 1^p + \delta^p = \delta^p + 1$ where the inequality holds by the concavity of $x^p$ for $p < 1$, and the claim holds by the monotonicity of $x^{1/p}$. We summarize the last two lemmas by the following straightforward corollary, which we will use.

**Corollary 4** *Consider a pair of machines $i$ and $i'$ such that $i < i'$. If $W_i > \frac{1}{\gamma(\varepsilon)} W_{i'}$ for some function $\gamma$ (such that $\gamma(\varepsilon) \geq \frac{1}{\varepsilon}$), then there is a function $\beta$ such that the ratio between the speeds $\max\{s_i, s_{i'}\}$ and $\min\{s_i, s_{i'}\}$ is bounded by $\beta(\varepsilon)$.*

**First rounding step.** In what follows we assume without loss of generality that the speeds are integer powers of $1+\varepsilon$. This assumption is justified by the observation that increasing the speed of each machine to the next value of the form $(1+\varepsilon)^j$ (for integer $j$) may decrease the completion time of this machine by a multiplicative factor of at most $1+\varepsilon$. Thus approximating the optimization problem with respect to the new speeds within a factor of $1+\varepsilon$ gives a $(1+\varepsilon)^{1+p}$ approximation to the original instance of the problem. Thus by scaling $\varepsilon$ accordingly, the assumption is justified. Moreover, we assume also that the sizes of all jobs are integer powers of $1+\varepsilon$. This assumption is justified by the observation that increasing the size of each job to the next value of the form $(1+\varepsilon)^j$ (for an integer $j$) may increase the completion time of each machine by a multiplicative factor of at most $1+\varepsilon$ and may not decrease it. Thus the following properties can be assumed.

**Assumption 5** *The speed of each machine as well as the size of each job is an integer power of $1+\varepsilon$.*

**Second rounding step.** Let $p_{\max} = \max_{j=1,2,\ldots,n} p_j$. We will let OPT denote a fixed optimal solution for the resulting instance after the second and final rounding step which we now define. For the minimization problem, we apply the following rounding (down) of the (rounded) processing times. If $p_j \leq \frac{\varepsilon p_{\max}}{n}$ then we round $p_j$ down to be zero and we remove all such jobs from the instance. Otherwise, we round $p_j$ down to the next integer multiple of $\mu = \frac{\varepsilon^2 p_{\max}}{n}$. Given a solution for the input after this second rounding step, we create a solution for the original instance, by assigning all the removed jobs to the machine in which a job of size $p_{\max}$ is assigned (breaking ties arbitrarily).

**Lemma 6** *The cost of every solution to the instance of the minimization problem after the second rounding step is no larger than its cost before this rounding step. Moreover, given a solution to the final instance, the cost of the resulting solution for the input after the first rounding step is at most $(1+\varepsilon)^{2p}$ times its cost for the new instance.*

**Proof.** Since we only round down processing times of jobs, the cost of the new solution cannot increase. Therefore, the first claim holds. As for the second claim we first consider the effect of returning the removed jobs to the solution. The work of the machine which receives these jobs increases by at most $\varepsilon p_{\max}$ and since it was previously at least $p_{\max}$, it increases by a multiplicative factor of at most $1+\varepsilon$.

Next, consider the effect of reverting the processing time of the jobs whose processing times were larger than $\frac{\varepsilon p_{\max}}{n}$ to their values in the instance after the first rounding step. Note that such a processing time of a job is increased by an additive factor of at most $\frac{\varepsilon^2 p_{\max}}{n}$ and hence by a multiplicative factor of at most $1+\varepsilon$. Therefore, the work of each machine is increased by a multiplicative factor of at most $1+\varepsilon$, and the claim follows. ∎

For the maximization problem, the second rounding step is defined as follows. In this case we let $\mu = \frac{\varepsilon^{\frac{1}{p}} \cdot p_{\max}}{n \cdot m^{\frac{1}{p}}}$, and we round the processing time of every job down to an integer multiple of $\mu$.

**Lemma 7** *The value of every solution to the new instance of the maximization problem after the second rounding step is not larger than its value before this rounding step and thus reverting jobs to their sizes after the first rounding step can only improve the performance. Moreover, consider an optimal solution SOL to the instance after the first rounding step. Denote by $\text{SOL}_{new}$, $\text{SOL}_{old}$ its objective function values in the new instance after the second rounding step, and before the second rounding step, respectively. Then, $\text{OPT} \geq \text{SOL}_{new} \geq (1-\varepsilon)\text{SOL}_{old}$.*

Note that after the second rounding step the size of any job is an integer multiple of $\mu$.

**Lemma 8** *Given an interval $[L, U]$, the number of distinct sizes of jobs is at most $\log_{1+\varepsilon} \frac{U}{L} + 2$.*

**Proof.** At the end of the first rounding step the number of distinct sizes of jobs in this interval is at most $\log_{1+\varepsilon} \frac{U}{L} + 1$, and this number may increase by at most 1 due to the second rounding step (in case where we round down a value which is slightly larger than $U$). ∎

## 3 Approximating the problem with a bounded weight ratio

In this section we consider the following variant of our (maximization or minimization) problem which is called BOUNDED RATIO (BR). The input to this problem consists of the following parts:
1. A set of $\ell \geq 1$ consecutive machines with speeds $s_i, s_{i+1}, \ldots, s_{i+\ell-1}$, for which we define

$$s_{\min} = \min\{s_i, s_{i+1}, \ldots, s_{i+\ell-1}\} \text{ and } s_{\max} = \max\{s_i, s_{i+1}, \ldots, s_{i+\ell-1}\},$$

and assume $\frac{s_{\max}}{s_{\min}} \leq \beta(\varepsilon)$. Recall that for the minimization problem we assume that the speeds are non-decreasing and thus $s_{\min} = s_i$ and $s_{\max} = s_{i+\ell-1}$, and for the maximization problem we assume that the speeds are non-increasing and thus $s_{\min} = s_{i+\ell-1}$ and $s_{\max} = s_i$.
2. A pair of values $\mu \leq \mathcal{W}_i \leq \mathcal{W}_{i+\ell-1}$ bounding the weights of machine $i$ and machine $i + \ell - 1$, respectively, such that $\frac{\mathcal{W}_{i+\ell-1}}{\mathcal{W}_i} \leq \gamma(\varepsilon)$. $\mathcal{W}_i, \mathcal{W}_{i+\ell-1}$ are (not necessarily positive) integer powers of $1 + \varepsilon$.
3. A value $A$ which is an integer multiple of $\mu$. For the minimization problem $A$ is an upper bound on the total size of jobs that the algorithm does not necessarily need to assign to any of these machines, whereas for the maximization problem the value of $A$ is a lower bound of the total size of jobs that the algorithm should not assign to any of these machines.
4. A set $L$ of *large* jobs $1, 2, \ldots, n$ each of size at least $\varepsilon \mathcal{W}_i$.
5. A value $B$ of the total size of existing *small* jobs. The small jobs can be assigned fractionally. The value $B$ is an integer multiple of $\mu$.

The goal is to schedule the large jobs and the small jobs on the $\ell$ machines such that the weight of each machine is at least $\mathcal{W}_i$ and at most $\mathcal{W}_{i+\ell-1}$. We allow an arbitrary subset of the jobs (out of the large jobs and small jobs) to remain unscheduled as long as in the minimization problem its total size is at most $A$, and in the maximization problem the total size of the unscheduled jobs must be at least $A$. We assume that such an assignment of the jobs is feasible or else the algorithm returns FALSE. The goal is to minimize or maximize the value $\sum_{j=i}^{i+\ell-1} C_j^p$ of the schedule, and the input to BR consists of $i$, $\ell$, $\mathcal{W}_i, \mathcal{W}_{i+\ell-1}, A, L$ and $B$. We will allow the algorithm (not the optimal solution) to use machines with weights in the interval $[\frac{\mathcal{W}_i}{1+3\varepsilon}, \mathcal{W}_{i+\ell-1} \cdot (1 + 3\varepsilon)]$.

**Remark 9** *By Corollary 4, the requirement that $\frac{s_{\max}}{s_{\min}} \leq \beta(\varepsilon)$ follows from $\frac{\mathcal{W}_{i+\ell-1}}{\mathcal{W}_i} \leq \gamma(\varepsilon)$.*

Next, the number of different sizes of large jobs is at most $\log_{1+\varepsilon} \frac{\mathcal{W}_{i+\ell-1}}{\varepsilon \mathcal{W}_i} + 2 \leq \log_{1+\varepsilon} \frac{\gamma(\varepsilon)}{\varepsilon} + 2$ and thus this number is a function of $\varepsilon$. Let $H$ be the set of different sizes of large jobs, and for each $h \in H$ we let $n_h$ be the number of jobs of size $h$. We define a class of machines to be machines with the same (rounded) speed. Since the speeds are integer powers of $1 + \varepsilon$ and the ratio between speeds satisfies $\frac{s_{\max}}{s_{\min}} \leq \beta(\varepsilon)$ we conclude that the number of non-empty machine classes, denoted as $\tau(\varepsilon)$, is at most $\log_{1+\varepsilon} \beta(\varepsilon) + 1$ that is a function of $\varepsilon$. We denote the non-empty machine classes in our problem by $M_1, \ldots, M_{\tau(\varepsilon)}$. For each machine class $M_k$, whose machines have a common speed of $\sigma_k$, we denote by $\nu(\sigma_k) = |M_k|$ the number of machines in $M_k$.

We define a *configuration $K$ of a machine* as a vector with the following components. The first $|H|$ components of $K$ define the number of large jobs of each size which we schedule on a machine with configuration $K$. For each $h \in H$, $n(h, K)$ denotes the number of jobs of size $h$ which are scheduled on a machine with this configuration. Each $n(h, K)$ is a non-negative integer which is at most $\frac{\mathcal{W}_{i+\ell-1}}{\varepsilon \mathcal{W}_i} \leq \frac{\gamma(\varepsilon)}{\varepsilon}$, that is a function of $\varepsilon$. The next component of $K$ is an integer power of $1+\varepsilon$ in the range $[\mathcal{W}_i, \mathcal{W}_{i+\ell-1}]$ denoted as $w(K)$. For the minimization problem, let $\tilde{w}(K) = \lfloor \frac{w(K)}{\mu} \rfloor \cdot \mu$. $\tilde{w}(K)$ is the maximum total size of jobs in configuration $K$. For the maximization problem, let $\tilde{w}(K) = \lceil \frac{w(K)}{\mu} \rceil \cdot \mu$. $\tilde{w}(K)$ is the minimum total size of jobs in configuration $K$. The number of options for this component is at most $\log_{1+\varepsilon} \frac{\mathcal{W}_{i+\ell-1}}{\mathcal{W}_i} + 1 \leq \log_{1+\varepsilon} \gamma(\varepsilon) + 1$, that is a function of $\varepsilon$. The last component is the machine speed, and we denote this component by $s(K)$. There are $\tau(\varepsilon)$ options for this last component. We conclude that the number of different configurations is a function of $\varepsilon$ and we can enumerate all of them in a constant time. We denote by $\mathcal{K}$ the set of all configurations. A configuration of a machine defines the number of large jobs of each size which are scheduled on such a machine, as well as the total size of small jobs which are scheduled fractionally on such a machine (which is the difference between $\tilde{w}(K)$ and the total size of the large jobs). Since the size of a job is an integer multiple of $\mu$, we require that the total size of small jobs which are scheduled on such a machine is an integer multiple of $\mu$ as well.

We define an integer program of fixed dimension to solve BR. The decision variables are for each configuration $K \in \mathcal{K}$, a variable $x_K$ counting the number of machines which are scheduled according to configuration $K$. We let $y_h$ be the number of large jobs of size $h \in H$ which remain unscheduled in our solution.

The following integer program is used for solving our minimization problem.

$$\min \quad \sum_{K \in \mathcal{K}} \left(\frac{w(K)}{s(K)}\right)^p \cdot x_K$$
$$\text{s.t.} \quad \sum_{K \in \mathcal{K}: s(K)=\sigma_k} x_K = \nu(\sigma_k) \quad \forall k=1,2,\ldots \tau(\varepsilon) \quad (1)$$
$$\sum_{K \in \mathcal{K}} n(h, K) \cdot x_K + y_h = n_h \quad \forall h \in H \quad (2)$$
$$\sum_{h \in H} h \cdot y_h - \sum_{K \in \mathcal{K}} x_K \cdot \left(\tilde{w}(K) - \sum_{h \in H} h \cdot n(h, K)\right) \leq A - B \quad (3)$$
$$\sum_{h \in H} h \cdot y_h \leq A \quad (4)$$
$$x_K, y_h \geq 0 \quad \forall K \in \mathcal{K}, \forall h \in H. (5)$$

The family of constraints (1) enforce that we use only $\nu(\sigma_k)$ machines with speed $\sigma_k$. The family of constraints (2) enforce that exactly $y_h$ jobs of size $h$ are unscheduled by our solution. Constraints (3) and (4) enforce the condition on the total size of jobs which are unscheduled. To see this last claim first note that a machine which is scheduled with configuration $K$ leaves a gap of size $\tilde{w}(K) - \sum_{h \in H} h \cdot n(h, K)$ for a possible scheduling of small jobs (since all jobs have sizes which are integer multiple of $\mu$). There are two cases. In the first case the total size of the gaps (of all the machines) is sufficient for scheduling all the small jobs that is $\sum_{K \in \mathcal{K}} x_K \cdot \left(\tilde{w}(K) - \sum_{h \in H} h \cdot n(h, K)\right) \geq B$. In this case, we assume without loss of generality that all the small jobs are scheduled. Thus in this case we only need to make sure that constraint (4) holds. This last constraint holds because the total size of the large jobs which are unscheduled is exactly $\sum_{h \in H} h \cdot y_h$, and thus such a feasible solution satisfies both constraints (3) and (4). In the other case there are small jobs which are not scheduled by the solution. Since we allow fractional scheduling of small jobs, their total size is exactly $B - \sum_{K \in \mathcal{K}} x_K \cdot \left(\tilde{w}(K) - \sum_{h \in H} h \cdot n(h, K)\right)$ which is positive. Thus the total size of the large jobs which may be unscheduled is at most $A - \left(B - \sum_{K \in \mathcal{K}} x_K \cdot \left(\tilde{w}(K) - \sum_{h \in H} h \cdot n(h, K)\right)\right)$ and thus constraint (3) holds, and by the assumption of this case, we conclude that constraint (4) holds as well.

Let $(x^*, y^*)$ be an optimal solution for the above integer program, and let $X^* = \sum_{K \in \mathcal{K}} \left(\frac{w(K)}{s(K)}\right)^p \cdot x_K^*$ be its objective function value.

**Claim 10** *Denote by $\text{OPT}_{br}$ the optimal solution for this bounded ratio minimization problem as well as the value of its objective function. We have $X^* \leq (1+\varepsilon)^p \cdot \text{OPT}_{br}$.*

The following integer program is used for solving our maximization problem.

$$\max \sum_{K \in \mathcal{K}} \left(\frac{w(K)}{s(K)}\right)^p \cdot x_K$$

$$\text{s.t.} \quad \sum_{K \in \mathcal{K}: s(K)=\sigma_k} x_K = \nu(\sigma_k) \quad \forall k = 1, 2, \ldots \tau(\varepsilon) \quad (6)$$

$$\sum_{K \in \mathcal{K}} n(h, K) \cdot x_K + y_h = n_h \quad \forall h \in H \quad (7)$$

$$\sum_{h \in H} h \cdot y_h - \sum_{K \in \mathcal{K}} x_K \cdot \left(\tilde{w}(K) - \sum_{h \in H} h \cdot n(h, K)\right) \geq A - B \quad (8)$$

$$\sum_{K \in \mathcal{K}} x_K \cdot \left(\tilde{w}(K) - \sum_{h \in H} h \cdot n(h, K)\right) \leq B \quad (9)$$

$$x_K, y_h \geq 0 \quad \forall K \in \mathcal{K}, \forall h \in H \quad (10)$$

The family of constraints (6) enforce that we use only $\nu(\sigma_k)$ machines with speed $\sigma_k$. The family of constraints (7) enforce that exactly $y_h$ jobs of size $h$ are unscheduled by our solution. Constraints (8) and (9) enforce the condition on the total size of jobs which are unscheduled. To see this last claim first we observe that a machine which is scheduled with configuration $K$ leaves a gap of size $\tilde{w}(K) - \sum_{h \in H} h \cdot n(h, K)$ which must be covered by small jobs. Therefore, the total size of small jobs should be sufficient to fill all these gaps. This enforces constraint (9). The remaining small jobs together with the unscheduled large jobs need to be of total size of at least $A$. Thus constraint (8) holds as well.

Let $(x^*, y^*)$ be an optimal solution for the above integer program, and let $X^* = \sum_{K \in \mathcal{K}} \left(\frac{w(K)}{s(K)}\right)^p \cdot x_K^*$ be its objective function value.

**Claim 11** *Denote by* $\text{OPT}_{br}$ *the optimal solution for this bounded ratio optimization problem as well as the value of its objective function. For the maximization problem* $X^* \geq \frac{\text{OPT}_{br}}{(1+\varepsilon)^p}$.

For each of our problems, we first solve the corresponding integer program. We show that given the solution $(x^*, y^*)$ to the integer program, we can construct a feasible solution for BR whose objective function value is at least as good as $X^*$. Then large jobs are assigned to the machines according to the configurations of the machines. Small jobs are assigned using (fractional) next-fit to the remaining gaps. In fractional next-fit we assign the jobs one by one until the current gap does not have a sufficient room for the next job, in which case we assign a fraction of the job to the current gap, so as to fill exactly the gap, and the remaining of the current job is assigned to the next gap (of the next machine). The work of a configuration $K$ is assumed to be $\tilde{w}(K)$. This procedure fills exactly all the gaps until one of the following two cases occurs. Either, there are no additional machines and there are still small jobs to be assigned, or there are no additional small jobs, but remaining gaps. By constraint (9) in the maximization problem only the second case may occur. In the minimization problem the first case does not cause any problem as it gives a feasible solution to problem BR since the remaining small jobs are counted towards $A$ in Constraint (3). In the second case for maximization problem, the unscheduled small jobs (or parts of these) are counted towards the total size of unscheduled jobs as well. This assignment algorithm is clearly a polynomial time algorithm. By the above claims it suffices to show that we can solve the integer programs in polynomial time.

**Claim 12** *The integer programs can be solved in strongly polynomial time.*

**Proposition 13** *Problem BR has an EPTAS.*

We next consider a variation of problem BR in which the small jobs which are scheduled on one of the $\ell$ machines need to be scheduled integrally. We call the resulting problem INTEGER BOUNDED RATIO (IBR). In order to obtain the EPTAS for IBR, we note that in our algorithm for BR, each machine receives at most two small jobs fractionally. For the maximization problem of IBR we simply remove the fractional parts. This decreases the completion time of each machine by at most $2\varepsilon \mathcal{W}_i$, and thus the completion time of each machine is decreased by a multiplicative factor of at most $1 - 2\varepsilon$. This gives an EPTAS for the maximization problem of IBR. For the minimization problem we assign (integrally) each small job to the first machine which gets a fraction of the job in the solution to BR. This may increase the completion time of a machine by at most $\varepsilon \mathcal{W}_i$. Therefore, the completion time of each machine increases by a multiplicative factor of at most $1+\varepsilon$ (it may decrease for some machines as well). Hence, in this case the total cost of the resulting solution to IBR is at most $(1+\varepsilon)^p$ times the cost of the solution to BR. Thus, this gives an EPTAS for the minimization problem of IBR as well, and the following result is established.

**Theorem 14** *Problem IBR has an EPTAS.*

## 4 Applying the shifting technique

In this section we use the shifting technique of Hochbaum and Maas [22]. We modify the total weight of a machine if it lies within *illegal intervals*. We will choose the set of illegal intervals (which imply a set of valid intervals) such that the following properties hold: First, the value of the objective function of an optimal solution using the modified weights will be close to its value according to the objective function of the original instance. Second, the ratio of weights of two values which are not separated by an illegal interval that is, belong to one valid interval will be bounded by a function $\gamma(\varepsilon)$. Finally, the ratio between the two extreme points of an illegal interval is at least $\frac{1}{\varepsilon}$. The set of illegal intervals is finite, and each such interval is bounded.

Given such a set of illegal intervals $S = \{(a_0, b_0), (a_1, b_1), \ldots, (a_r, b_r)\}$ where $b_\ell < a_{\ell+1}$ for all $0 \leq \ell \leq r-1$ and $a_0 \geq \min_j p_j$ and $b_r \leq \sum_{j=1}^n p_j$. We consider a schedule and assume that when a machine $i$ gets an allocation of a total weight $W_i$, it will contribute $\left(\frac{g(W_i)}{s_i}\right)^p$ to the objective function. We next define the function $g$.

For the maximization problem where $p < 1$, we define $g(x) = x$ if $x \notin (a_\ell, b_\ell)$ for all $\ell$, and otherwise $g(x) = 0$. For the minimization problem where $p > 1$, we define $g(x) = x$ if $x \notin (a_\ell, b_\ell)$ for all $\ell$, and otherwise if $x \in (a_\ell, b_\ell)$ we let $g(x) = 2b_\ell$. By using the value of $g$ instead of the total weight, the objective function value of any feasible solution may become worse (larger for the minimization problem and smaller for the maximization problem). We denote by $p_{\min}$ the minimum size of a job in the instance.

Our algorithm will choose the best outcome among a constant number of iterations. In each such iteration we will use a different set of illegal intervals. Let $\rho = \frac{1}{\varepsilon^{2\lceil p\rceil+1}} \geq \frac{1}{\varepsilon^3}$ (where equality holds for $p < 1$), then for $\eta = 0, 1, \ldots, \rho - 1$, in iteration $\eta$ we will use the following set of illegal intervals: $(a_\ell, b_\ell) = (p_{\min} \cdot \left(\frac{1}{\varepsilon}\right)^{\eta+\ell\rho}, p_{\min} \cdot \left(\frac{1}{\varepsilon}\right)^{\eta+\ell\rho+1})$, for the non-negative values of $\ell$ such that $b_\ell \leq \sum_{j=1}^n p_j$. The number of non-negative powers of $\frac{1}{\varepsilon}$ for which we define an interval in all iterations is at most

$$\log_{\frac{1}{\varepsilon}} \frac{np_{\max}}{p_{\min}} + 1 \leq \log_{\frac{1}{\varepsilon}} \frac{np_{\max}}{\mu} + 1 \leq \log_{\frac{1}{\varepsilon}} \frac{n^2 \cdot m^{1/p}}{\varepsilon^{1/p+1}} + 1,$$

and this is a polynomial in $n, m$ and $\frac{1}{\varepsilon}$. We denote by $g_\eta$ the function $g$ in the $\eta$-th iteration, and we denote by $S_\eta$ the set of illegal intervals of iteration $\eta$.

We next show that there is a value of $\eta$ such that the objective function value of the optimal solution to the problem with the modified weights is within a factor of $1+\varepsilon$ of the objective function value of an optimal solution to the original instance.

**Lemma 15** *Denote by* $\text{OPT}_\eta$ *the objective function value of an optimal solution with respect to the modified weights* $g_\eta$. *Then, for the maximization problem there is a value of $\eta$ such that* $\text{OPT} \geq \text{OPT}_\eta \geq \text{OPT} \cdot (1-\varepsilon^3) \geq \frac{\text{OPT}}{1+\varepsilon}$, *and for the minimization problem there is a value of $\eta$ such that* $\text{OPT} \leq \text{OPT}_\eta \leq \text{OPT}(1+\varepsilon)$.

We next show that given an instance where the value or cost of a solution is computed using $g_\eta$, we can restrict ourselves to solutions which (almost) do not use weights from $S_\eta$. More precisely we show the following claim.

**Lemma 16** *Consider the optimal solution $\text{OPT}_\eta$ for the instance with the modified weights $g_\eta$. Without loss of generality, there is at most one machine $i$ whose weight $W_i$ belongs to one of the intervals of $S_\eta$.*

Note that the number of machines of weight zero can be arbitrary.

**Proof.** Denote by $M_\eta$ the set of machines whose weight belong to an interval from $S_\eta$. Consider the maximization problem. If the claim does not hold, then by assumption $|M_\eta| \geq 2$. We modify the solution so that all the jobs which were assigned to one of the machines in $M_\eta$ are assigned to the maximum index machine of $M_\eta$. The machines of $M_\eta$ did not contribute a positive value to the objective function in $\text{OPT}_\eta$ (since the modified objective function is used), and hence this modification did not hurt the optimality of $\text{OPT}_\eta$. The claim holds for the new solution.

Consider the minimization problem. We repeat the following modification of $\text{OPT}_\eta$ as long as $|M_\eta| \geq 2$. Let $i \in M_\eta$ be a most loaded machine in $\text{OPT}_\eta$ among the machines of $M_\eta$. Let $i' \neq i$ be another member of $M_\eta$. Then, the weight of $i'$ in $\text{OPT}_\eta$ is at most $W_i$. We modify $\text{OPT}_\eta$ by assigning to machine $i$ all jobs which were assigned previously to either $i$ or $i'$. This modification will at most double the weight of $i$, and hence the solution remains optimal (either $i$ remains in $M_\eta$ and in this case it pays the same as in the original solution and $i'$ pays nothing, or $i$ is removed from $M_\eta$ and in this case its work is at most double the infimum point of the following allowed interval so it now pays at most its payment in the original solution, and $i'$ pays nothing). Repeating the process decreases the cardinality of $M_\eta$ ($i'$ is removed from $M_\eta$ after its weight becomes zero). ∎

Next, given such a value of $\eta$, we can guess the parity of the index $\ell$ (of the interval $(a_\ell, b_\ell) \in S_\eta$ which contains a weight of a machine in $\text{OPT}_\eta$ if it exists). By the previous lemma there is at most one such value of $\ell$. Given such a guess, we allow the use of intervals of the same parity as $\ell$. That is, if $\ell$ is an odd number, we remove from $S_\eta$ the intervals of the form $(a_{2i-1}, b_{2i-1})$, and if $\ell$ is an even number, we remove from $S_\eta$ the intervals of the form $(a_{2i}, b_{2i})$. We denote the resulting set of illegal intervals by $S_{\eta,\phi}$ where $\phi \in \{\text{odd}, \text{even}\}$. The number of possibilities for this guess (of $\eta$ and $\phi$) is polynomial in $\frac{1}{\varepsilon}$. Hence, we can assume that the set $S_{\eta,\phi}$ satisfies that $\text{OPT}_\eta$ does not have a machine whose weight belongs to an interval from $S_{\eta,\phi}$. Moreover, the following observation holds.

**Observation 17** *The ratio between two weights $W < W'$ which are not separated by an interval from $S_{\eta,\phi}$ is bounded by a function $\gamma(\varepsilon) = \frac{1}{\varepsilon^{2\rho-1}}$ (i.e., $\frac{W'}{W} \leq \gamma(\varepsilon)$), and the ratio between $W'$ and $W$ if they are separated by an interval from $S_{\eta,\phi}$ is at least $\frac{1}{\varepsilon}$.*

Using this set of intervals $S_{\eta,\phi}$, we can evaluate all the solutions which satisfy the property that they do not use weights from the set $S_{\eta,\phi}$, according to the weights and not the modified weights. This cannot harm the quality of the solution with respect to the modified objective functions. We conclude that using such solutions which do not use a weight from $S_{\eta,\phi}$ does not hurt the approximation algorithm too much. That is, we establish the following lemma.

**Lemma 18** *There is a value of $\eta$ and $\phi$, such that $\text{OPT}_\eta$ does not use a weight which belong to an interval of $S_{\eta,\phi}$, and such that the objective function value of $\text{OPT}_\eta$ (with respect to the original objective function) in terms of the weights (i.e., as a solution to the original instance) is within a factor of $1+\varepsilon$ of $\text{OPT}$.*

## 5 Dynamic programming to approximate $\text{OPT}_\eta$

Given fixed values of $\eta$ and $\phi$, the set of illegal weights $S_{\eta,\phi}$ leaves a set of valid intervals whose possible weights denoted by $\Omega = \{[\omega_0, \omega_1], [\omega_2, \omega_3], \ldots, [\omega_{r'-1}, \omega_{r'}]\}$ where the sequence $\omega_i$ is monotone increasing, $\omega_0 \geq p_{\min}$, and $r' \leq r$. Recall that some machines may get a zero weight in an optimal solution, but we can guess their number and remove the set of these machines (of lowest indices) from the instance. Hence, without loss of generality it suffices to consider solutions which assign at least one job to each machine, and therefore the minimum weight of a machine is at least $\omega_0$.

By Observation 17, we conclude that for any value of $\xi$, $\frac{\omega_{2\xi}}{\omega_{2\xi-1}} = \frac{1}{\varepsilon}$ and $\frac{\omega_{2\xi+1}}{\omega_{2\xi}} \leq \gamma(\varepsilon)$. We define a linear order on the intervals of $\Omega$ saying that an interval $[\omega_\xi, \omega_{\xi+1}]$ is smaller than $[\omega_{\xi'}, \omega_{\xi'+1}]$ if $\xi < \xi'$, and an interval $[\omega_\xi, \omega_{\xi+1}]$ is at most an interval $[\omega_{\xi'}, \omega_{\xi'+1}]$ if $\xi \leq \xi'$. We next describe an allocation of jobs to intervals of $\Omega$. A job $j$ of size $p_j$ is associated with an interval $[\omega_{\xi-1}, \omega_\xi]$ if $p_j \leq \omega_\xi$ and $p_j > \omega_{\xi-2}$ where we use the convention $\omega_{-1} = 0$. We define an assignment of intervals of (consecutive) machines to intervals of $\Omega$ in the following sense. An interval of machines $[i, i']$ with parameters $A, B$ is assigned to an interval $[\omega_\xi, \omega_{\xi+1}] \in \Omega$ if the following four conditions hold: 1) the weight of each machine $\tilde{i} \in [i, i'-1]$ is in the interval $[\omega_\xi, \omega_{\xi+1}]$; 2) no other machine has weight in this interval; 3) the total size of jobs associated with smaller intervals and are scheduled by a machine of index at least $i$ is $B$; and similarly 4) the total size of jobs associated with intervals at most $[\omega_\xi, \omega_{\xi+1}]$ and are scheduled by a machine of index at least $i'$ is $A$. Here $A$ and $B$ are integer multiples of $\mu$. We say that $A$ and $B$ are the parameters of the interval of machines $[i, i']$.

**Claim 19** *The number of possibilities of a machine interval and a pair of parameters is polynomial in the input size.*

Note that given an interval of machines $[i, i']$ and values $A, B$ we get an instance of the IBR problem for which we presented an EPTAS (this IBR instance has an empty set of machines if $i = i'$ and otherwise at least one machine). Thus our scheme applies this EPTAS for each possibility of a machine interval $[i, i']$ such that $i \leq i'$ corresponding to weight interval $[\omega_\xi, \omega_{\xi+1}]$ with values $A, B$. We denote by $IBR_{eptas}(i, i'-1, \xi, A, B)$ the solution returned by the EPTAS for the IBR instance as well as its objective function value. If $i = i'$ and $A = B$, then $IBR_{eptas}(i, i'-1, \xi, A, B) = 0$, and if $i = i'$ and $A \neq B$ then $IBR_{eptas}(i, i'-1, \xi, A, B)$ returns FALSE. If the returned output is FALSE, then the value is $\infty$ for the minimization problem and $-\infty$ for the maximization problem.

To find the approximated solution for the minimization problem we find a shortest (minimum cost) path in the following layered graph $G$, and to find the approximated solution for the maximization problem we find a longest (maximum cost) path in this graph $G$ (since $G$ is a layered graph, it is acyclic and hence both the shortest path problem and the longest path problem are solvable in linear time). We have a layer for each value of $\xi$ such that $[\omega_\xi, \omega_{\xi+1}] \in \Omega$ (that is, we will have a layer for every even value of $\xi$). Each such layer corresponding to $\xi$ has a vertex for each machine (and one additional dummy machine of index $m+1$), and each possibility for the value of $A$ (i.e., for each integer multiple of $\mu$). Given a vertex $(i, b)$ in layer $\xi$ and a vertex $(i', a)$ in layer $\xi + 2$, there is an arc from the former vertex to the later vertex if $i \leq i'$, and the cost associated with such an arc is $IBR_{eptas}(i, i'-1, \xi, a, b)$. The construction of $G$ takes polynomial time, and it has a polynomial size.

We next find a shortest or longest path in $G$ from the vertex $(1, 0)$ of the layer with index $0$ to the vertex $(m+1, 0)$ in the last layer. We schedule the jobs according to the path which we found. That is, if the path uses the arc from $(i, b)$ of layer $\xi$ to $(i', a)$ of layer $\xi + 2$ where $i \leq i'$, then we schedule large and small jobs as defined by the solution $IBR_{eptas}(i, i'-1, \xi, a, b)$. The total size of jobs which are associated with the interval $[\omega_\xi, \omega_{\xi+1}]$ or smaller intervals, and are not scheduled to machines of index at most $i'-1$ is indeed at most $a$ for the minimization problem and at least $a$ for the maximization problem. The EPTAS for IBR may use at most one small job which exceeds the total of $B$. This job is scheduled in the case of minimization to the machine $i'-1$ and discarded in the case of maximization. In addition, since we use only some values for $w(K)$, in the case of minimization we have too large room for small jobs, and in the case of maximization too small room for small jobs. Thus, for the minimization problem, it can be the case that too many jobs are scheduled and as a result no small available jobs remain. In this case, the algorithm moves to deal with the next arc of the path (or stops if this is the last arc). Similarly, for the maximization problem, since discarded jobs remain available, there are always sufficiently many small jobs to cover all the configurations (and some small jobs may remain unassign upon the termination (at the end of the path). Note that every job is scheduled by one of the solutions corresponding to this path, and hence we obtained a feasible solution whose objective function value is the total cost of the arcs in the path.

We next note that $\text{OPT}_\eta$ also corresponds to a path in $G$ in the following sense. If $\text{OPT}_\eta$ uses machines $i, i+1, \ldots, i'-1$ with weight in the interval $[\omega_\xi, \omega_{\xi+1}]$, the total size of the jobs allocated to smaller intervals and are not scheduled by $\text{OPT}_\eta$ to machines with index at most $i-1$ is $b$, and the total size of the jobs allocated to intervals at most this interval and are not scheduled by $\text{OPT}_\eta$ to machines with index at most $i'-1$ is $a$, then we say that the arc from $(i, b)$ of layer $\xi$ to $(i', a)$ of layer $\xi + 2$ belongs to the path associated with $\text{OPT}_\eta$. Since $\text{OPT}_\eta$ is a feasible solution, this set of arcs can be augmented to form a path by adding zero cost arcs (from $(i, b)$ of layer $\xi$ to $(i, b)$ of layer $\xi + 2$). Moreover, the cost of $\text{OPT}_\eta$ can be also partitioned into its arcs, by assigning the cost (according to $\text{OPT}_\eta$) of the machines $i, i+1, \ldots, i'-1$ to the arc from $(i, b)$ of layer $\xi$ to $(i', a)$ of layer $\xi + 2$. By the correctness of the EPTAS to IBR we conclude that the cost assigned to such an arc is within a multiplicative factor of $1+\varepsilon$ of the cost of the arc. Hence, the cost of this path in $G$ is within a factor of $1+\varepsilon$ of the objective function value of $\text{OPT}_\eta$. Since we use the optimal path, we are not worse than this path of $\text{OPT}_\eta$, and thus we established the following result.

**Theorem 20** *Both the problem of minimizing $\sum_{i=1}^m C_i^p$ and the problem of maximizing $\sum_{i=1}^m C_i^p$ for real finite values of $p$ have efficient polynomial time approximation schemes.*

# References

[1] N. Alon, Y. Azar, G. J. Woeginger, and T. Yadid. Approximation schemes for scheduling. In *SODA'97*, 493–500.

[2] N. Alon, Y. Azar, G. J. Woeginger, and T. Yadid. Approximation schemes for scheduling on parallel machines. *Journal of Scheduling*, 1(1):55–66, 1998.

[3] A. Avidor, Y. Azar, and J. Sgall. Ancient and new algorithms for load balancing in the $L_p$ norm. *Algorithmica*, 29(3):422–441, 2001.

[4] B. Awerbuch, Y. Azar, E. F. Grove, M.-Y. Kao, P. Krishnan, and J. S. Vitter. Load balancing in the $l_p$ norm. In *FOCS'95*, 383–391.

[5] Y. Azar and L. Epstein. Approximation schemes for covering and scheduling on related machines. In *APPROX'98*, 39–47.

[6] Y. Azar, L. Epstein, Y. Richter, and G. J. Woeginger. All-norm approximation algorithms. *Journal of Algorithms*, 52(2):120–133, 2004.

[7] N. Bansal and K. R. Pruhs. Server scheduling to balance priorities, fairness, and average quality of service. *SIAM Journal on Computing*, 39(7):3311–3335, 2010.

[8] N. Bansal and M. Sviridenko. The Santa Claus problem. In *STOC'06*, 31–40.

[9] I. Caragiannis. Better bounds for online load balancing on unrelated machines. In *SODA'08*, 972–981.

[10] M. Cesati and L. Trevisan. On the efficiency of polynomial time approximation schemes. *Information Processing Letters*, 64(4):165–171, 1997.

[11] A. K. Chandra and C. K. Wong. Worst-case analysis of a placement algorithm related to storage allocation. *SIAM Journal on Computing*, 4(3):249–263, 1975.

[12] R. A. Cody and E. G. Coffman Jr. Record allocation for minimizing expected retrieval costs on drum-like storage devices. *Journal of the ACM*, 23(1):103–115, 1976.

[13] R. G. Downey and M. R. Fellows. *Parameterized Complexity*. Springer-Verlag, Berlin, 1999.

[14] P. Efraimidis and P. G. Spirakis. Approximation schemes for scheduling and covering on unrelated machines. *Theoretical Computer Science*, 359(1-3):400–417, 2006.

[15] L. Epstein and J. Sgall. Approximation schemes for scheduling on uniformly related and identical parallel machines. *Algorithmica*, 39(1):43–57, 2004.

[16] L. Epstein and R. van Stee. Maximizing the minimum load for selfish agents. *Theoretical Computer Science*, 411(1):44–57, 2010.

[17] J. Flum and M. Grohe. *Parameterized Complexity Theory*. Springer-Verlag, Berlin, 2006.

[18] D. K. Friesen and B. L. Deuermeyer. Analysis of greedy solutions for a replacement part sequencing problem. *Mathematics of Operations Research*, 6(1):74–87, 1981.

[19] T. Gonzalez, O. H. Ibarra, and S. Sahni. Bounds for LPT schedules on uniform processors. *SIAM Journal on Computing*, 6(1):155–166, 1977.

[20] R. L. Graham. Bounds for certain multiprocessing anomalies. *Bell Sys. Tec. J.*, 45(9):1563–1581, 1966.

[21] D. S. Hochbaum. Various notions of approximations: Good, better, best and more. In D. S. Hochbaum, editor, *Approximation algorithms*. PWS Publishing Company, 1997.

[22] D. S. Hochbaum and W. Maass. Approximation schemes for covering and packing problems in image processing and VLSI. *Journal of the ACM*, 32(1):130–136, 1985.

[23] D. S. Hochbaum and D. B. Shmoys. Using dual approximation algorithms for scheduling problems: theoretical and practical results. *Journal of the ACM*, 34(1):144–162, 1987.

[24] D. S. Hochbaum and D. B. Shmoys. A polynomial approximation scheme for scheduling on uniform processors: Using the dual approximation approach. *SIAM J. on Computing*, 17(3):539–551, 1988.

[25] E. Horowitz and S. Sahni. Exact and approximate algorithms for scheduling nonidentical processors. *Journal of the ACM*, 23(2):317–327, 1976.

[26] K. Jansen. An EPTAS for scheduling jobs on uniform processors: Using an MILP relaxation with a constant number of integral variables. *SIAM Journal on Discrete Mathematics*, 24(2):457–485, 2010.

[27] H. W. Lenstra Jr. Integer programming with a fixed number of variables. *Mathematics of Operations Research*, 8(4):538–548, 1983.

[28] J. Y. T. Leung and W. D. Wei. Tighter bounds on a heuristic for a partition problem. *Information Processing Letters*, 56(1):51–57, 1995.

[29] D. Marx. Parameterized complexity and approximation algorithms. *The Comp. J.*, 51(1):60–78, 2008.

[30] G. J. Woeginger. A polynomial time approximation scheme for maximizing the minimum machine completion time. *Operations Research Letters*, 20(4):149–154, 1997.

# A  Omitted proofs

## A.1  Proof of Lemma 2

Assume by contradiction that $W_i > \delta W_{i'}$. The solution pays at least $\left(\frac{W_i}{s_i}\right)^p$ for this pair of machines. We move all the jobs which were scheduled on machine $i$ to machine $i'$. It suffices to show that the completion time of machine $i'$ in this new solution is smaller than $\frac{W_i}{s_i}$. This last claim holds because the completion time of $i'$ in the new solution is $\frac{W_i+W_{i'}}{s_{i'}} < \frac{2W_i}{\delta s_{i'}} \leq \frac{2W_i}{\delta s_i} \cdot \alpha(\delta) = \frac{W_i}{s_i}$ where the first inequality holds because $W_{i'} \leq \frac{W_i}{\delta}$ and $\delta < 1$, the second inequality holds because $s_i \leq \alpha(\delta) \cdot s_{i'}$, and the equality holds by the definition of $\alpha(\delta)$.

## A.2  Proof of Lemma 3

Assume by contradiction that the claim does not hold, that is, there exist such machines $i$ and $i'$ with $W_i > \delta W_{i'}$. We compare the current solution with a new solution that schedules all the jobs (which were previously scheduled on either $i$ or $i'$) to machine $i'$. It suffices to show that this new solution is better, that is, that the following inequality holds: $\left(\frac{W_i+W_{i'}}{s_{i'}}\right)^p > \left(\frac{W_i}{s_i}\right)^p + \left(\frac{W_{i'}}{s_{i'}}\right)^p$. We first note that $\left(\frac{W_i}{s_i}\right)^p \leq \left(\frac{W_{i'}}{s_{i'}}\right)^p \cdot ((1+\delta)^p - 1)$. This claim holds because $W_i \leq W_{i'}$ and $s_{i'} \leq \alpha(\delta) \cdot s_i$, and therefore $\left(\frac{W_i}{s_i}\right)^p \leq \left(\frac{W_{i'}}{s_{i'}}\right)^p \cdot (\alpha(\delta))^p$, and the claim holds by the definition of $\alpha(\delta)$. Therefore, to get a contradiction it suffices to show that $\left(\frac{W_i+W_{i'}}{s_{i'}}\right)^p > \left(\frac{W_{i'}}{s_{i'}}\right)^p \cdot (1+\delta)^p$, which is equivalent to $(W_i + W_{i'})^p > (W_{i'} \cdot (1+\delta))^p$ which holds using $W_i > \delta W_{i'}$, since $x^p$ is a strictly monotone increasing function of $x$ (for $0 \leq x < \infty$ and any $p > 0$).

## A.3  Proof of Lemma 7

The first claim holds because we round down the size of each job. We prove the second claim. Consider a machine $i$ whose work in SOL is denoted by $W_i$. If $W_i \leq n\mu$ then its new work is at least $0$. Otherwise, its new work is at least $W_i - n\mu$. Consider the case where $W_i > n\mu$. Then, the contribution of machine $i$ to $\text{SOL}_{new}$ is at least $\left(\frac{W_i-n\mu}{s_i}\right)^p \geq \left(\frac{W_i}{s_i}\right)^p - \left(\frac{n\mu}{s_i}\right)^p = \left(\frac{W_i}{s_i}\right)^p - \frac{\varepsilon p_{\max}^p}{m \cdot s_i^p} \geq \left(\frac{W_i}{s_i}\right)^p - \frac{\varepsilon p_{\max}^p}{m \cdot s_m^p}$ where the first inequality holds by concavity of $x^p$ for $p < 1$, and the last one holds using $s_i \geq s_m$.

Consider the case where $W_i \leq n\mu$. Then, the contribution of machine $i$ to $\text{SOL}_{new}$ is at least $0 \geq \left(\frac{W_i}{s_i}\right)^p - \left(\frac{n\mu}{s_i}\right)^p \geq \left(\frac{W_i}{s_i}\right)^p - \frac{\varepsilon p_{\max}^p}{m \cdot s_m^p}$, where the second inequality holds similarly to the previous case. Thus, the decrease of the contribution of each machine towards the objective function value of SOL, is at most $\frac{\varepsilon p_{\max}^p}{m \cdot s_m^p} \leq \frac{\varepsilon \text{SOL}_{old}}{m}$, where the inequality holds because $\text{SOL}_{old} \geq \left(\frac{p_{\max}}{s_m}\right)^p$ which is the value of a solution that places the largest job on the slowest machine, and ignores all the other jobs.

## A.4  Proof of Claim 10

First, note that $\text{OPT}_{br}$ induces a feasible solution for the integer program. To get this induced solution we apply the following rounding for each machine in the optimal solution: for each machine we define a configuration $K$ by first counting the number of large jobs of each size, and it remains to define $w(K)$. $w(K)$ is a rounded up value of the total weight of the jobs assigned to the machine to the closest integer power of $1+\varepsilon$. We next argue that the total work of this machine in $\text{OPT}_{br}$ is at most $\tilde{w}(K)$. This is so because otherwise, a machine with total work of $\theta$ in $\text{OPT}_{br}$ has a configuration $K$ with $w(K) \geq \theta$ which is an integer power of $1+\varepsilon$, and $\tilde{w}(K) = \theta'$ such that $\theta' < \theta$. However, since $\theta$ is an integer multiple of $\mu$ we have $\theta = \mu \cdot \lfloor \frac{\theta}{\mu} \rfloor \leq \mu \cdot \lfloor \frac{w(K)}{\mu} \rfloor = \theta'$, that is a contradiction. The total size of the gaps for scheduling the small jobs does not decrease, and thus the total size of the unscheduled jobs is at most $A$. This results in a feasible solution to the integer program. The value of the objective function of this rounded solution is at most $(1+\varepsilon)^p \cdot \text{OPT}_{br}$, and at least $X^*$, and thus the claim holds.

## A.5 Proof of Claim 11

First, note that $\text{OPT}_{br}$ induces a feasible solution for the integer program. To get this induced solution we apply the following rounding for each machine in the optimal solution: for each machine we define a configuration $K$ by first counting the number of large jobs of each size, and it remains to define $w(K)$. $w(K)$ is a rounded down value of the total weight of the jobs assigned to the machine to the closest integer power of $1+\varepsilon$. We next argue that the total work of this machine in $\text{OPT}_{br}$ is at least $\tilde{w}(K)$. This is so because otherwise, a machine with total work of $\theta$ in $\text{OPT}_{br}$ has a configuration $K$ with $w(K) \leq \theta$ which is an integer power of $1+\varepsilon$, and $\tilde{w}(K) = \theta'$ such that $\theta' > \theta$. However, since $\theta$ is an integer multiple of $\mu$ we have $\theta = \mu \cdot \lceil \frac{\theta}{\mu} \rceil \geq \mu \cdot \lceil \frac{w(K)}{\mu} \rceil = \theta'$, that is a contradiction. The total size of the gaps for scheduling the small jobs does not increase, and thus there is sufficient total size of small jobs to fill in all the gaps, and the total size of the unscheduled jobs is at least $A$. This results a feasible solution to the integer program. The value of the objective function of this rounded solution is at least $\frac{\text{OPT}_{br}}{(1+\varepsilon)^p}$, and at most $X^*$, and thus the claim holds.

## A.6 Proof of Claim 12

First, the construction of the integer programs takes polynomial time since the set of all configurations can be enumerated in polynomial time (using the fact that $\mathcal{K}$ has at most a constant number of configurations which is upper bounded by a function of $\varepsilon$). Next, we observe that the dimension of each of these programs (the number of variables) is $|\mathcal{K}| + |H|$, and as explained above, both $|\mathcal{K}|$ and $|H|$ are bounded by a function of $\frac{1}{\varepsilon}$. Thus the integer program has a fixed dimension, and we can use the polynomial time algorithms for solving such a problem. The number of constraints (beside the non-negativity constraints) is $\tau(\varepsilon) + |H| + 2$ which is again bounded by a function of $\frac{1}{\varepsilon}$. Therefore, using Lenstra's algorithm [27] or one of its improvements, give a polynomial time algorithm for solving (exactly) the integer programs (recall that the time complexity of solving an integer program of dimension $d$ is $f(d) \cdot poly$ where $f$ is an exponential function of the dimension, and $poly$ is a polynomial in the binary encoding of the program). To obtain a strongly polynomial time we use the following observations. First, the coefficients in the objective function are integer powers of $1+\varepsilon$ and can be scaled to be at most a function of $\varepsilon$ (since the ratios $\frac{s_{\max}}{s_{\min}}$ and $\frac{\mathcal{W}_{i+\ell-1}}{\mathcal{W}_i}$ are bounded). Next, we scale constraints (3), (4), (8), and (9) by dividing the constraints by the factor $\mu$. In the resulting constraint matrices and right hand sides, all the coefficients are strongly polynomial (i.e., do not depend on the magnitude of the numbers in the instance).

## A.7 Proof of Lemma 15

Fix an optimal solution $\text{OPT}$ to the original instance, and associate the value of each machine to a corresponding value of $\eta$ as follows. Assume that a machine $i$ has weight $W_i$, and assume that $W_i \in \left( p_{\min} \cdot \left( \frac{1}{\varepsilon} \right)^{\eta_0 + \ell\rho}, p_{\min} \cdot \left( \frac{1}{\varepsilon} \right)^{\eta_0 + \ell\rho + 1} \right)$, then we associate $\left( \frac{W_i}{s_i} \right)^p$ with this value $\eta_0$ of $\eta$. Note that for a machine $i$ there is at most one such corresponding value of $\eta$. If there is no such corresponding value of $\eta$ then we associate $\left( \frac{W_i}{s_i} \right)^p$ with an arbitrary value of $\eta$. By the pigeonhole principle, there is a value $\eta_0$ of $\eta$ which is associated with at most $\frac{\text{OPT}}{\rho}$.

Consider first the maximization problem. Using the value $\eta_0$ of $\eta$, OPT is a feasible solution to the problem with the modified value function $g_{\eta_0}$ whose objective function value is at least $\text{OPT} \cdot (1 - \frac{1}{\rho}) = \text{OPT}(1 - \varepsilon^3)$, and thus the claim holds.

Next, consider the minimization problem. Using the value $\eta_0$ of $\eta$, OPT is a feasible solution to the problem with the modified cost $g_{\eta_0}$ whose objective function value is at most $\text{OPT} + \frac{\text{OPT}}{\rho} \cdot \left(\frac{2}{\varepsilon}\right)^p \leq \text{OPT} + \frac{\text{OPT}}{\rho} \cdot \frac{1}{\varepsilon^{2p}} \leq \text{OPT}(1+\varepsilon)$, where the first inequality holds because the modified weight of a machine whose corresponding value of $\eta$ is $\eta_0$ is at most $\frac{2}{\varepsilon}$ times its weight. Thus, the claim holds.

## A.8 Proof of Claim 19

The number of pairs of machines $i$ and $i'$ is at most $m^2$, and the number of possibilities for the value $A$ (and similarly for $B$) is at most $\frac{n \cdot p_{\max}}{\mu} \leq \frac{n^2 \cdot m^{1/p}}{\varepsilon^{1/p+1}}$. Thus, the claim holds.